\documentclass[aps,prl,reprint,superscriptaddress,twocolumn]{revtex4}
\usepackage[dvips]{graphicx}
\usepackage{dcolumn}
\usepackage{bm}
\usepackage{amsmath}

\begin{document}

\title{Characterization of aluminum oxide tunnel barriers by combining transport measurements and transmission electron microscope imaging}

\author{T. Aref}
\affiliation{Low Temperature Laboratory (OVLL), Aalto University School of Science, P.O.~Box 13500, 00076 Aalto, Finland}
\altaffiliation{Current address: Microtechnology and Nanoscience, MC2, Chalmers University of Technology, SE-412 96 G\"oteborg, Sweden}

\author{A. Averin}
\affiliation{Low Temperature Laboratory (OVLL), Aalto University School of Science, P.O.~Box 13500, 00076 Aalto, Finland}

\author{S. van Dijken}
\affiliation{NanoSpin, Department of Applied Physics, Aalto University School of Science, P.O. Box 15100, FI-00076 Aalto, Finland}

\author{A. Ferring}
\affiliation{Kirchhoff-Institute for Physics, Heidelberg University, Im Neuenheimer Feld 227, D-69120 Heidelberg, Germany}

\author{M. Koberidze}
\affiliation{COMP/Department of Applied Physics, Aalto University School of Science, P.O. Box 11100, FI-00076 Aalto, Espoo, Finland}

\author{V. F. Maisi}
\email{ville.maisi@gmail.com}
\affiliation{Low Temperature Laboratory (OVLL), Aalto University School of Science, P.O.~Box 13500, 00076 Aalto, Finland}
\affiliation{Centre for Metrology and Accreditation (MIKES), P.O. Box 9, 02151 Espoo, Finland}
\altaffiliation{Current address: Solid State Physics Laboratory, ETH Zurich, CH-8093 Zurich, Switzerland}

\author{H. Nguyen}
\affiliation{Low Temperature Laboratory (OVLL), Aalto University School of Science, P.O.~Box 13500, 00076 Aalto, Finland}

\author{R. M. Nieminen}
\affiliation{COMP/Department of Applied Physics, Aalto University School of Science, P.O. Box 11100, FI-00076 Aalto, Espoo, Finland}

\author{J. P. Pekola}
\affiliation{Low Temperature Laboratory (OVLL), Aalto University School of Science, P.O.~Box 13500, 00076 Aalto, Finland}

\author{L. D. Yao}
\affiliation{NanoSpin, Department of Applied Physics, Aalto University School of Science, P.O. Box 15100, FI-00076 Aalto, Finland}

\begin{abstract} 
We present two approaches for studying the uniformity of a tunnel barrier. The first approach is based on measuring single-electron and two-electron tunneling in a hybrid single-electron transistor. Our measurements indicate that the effective area of a conduction channel is about one order of magnitude larger than predicted by theoretical calculations. With the second method, transmission electron microscopy, we demonstrate that variations in the barrier thickness are a plausible explanation for the larger effective area and an enhancement of higher order tunneling processes.
\end{abstract}

\maketitle

Tunnel junctions are used in metallic single-electron devices such as single-electron sources~\cite{Pekola2013}, superconducting qubits~\cite{Nakamura1999,Martinis2005,Paik2011,Abdumalikov2013} and electronic coolers~\cite{Nahum1994,Leivo1996,Clark2005}. An essential part of a tunnel junction is the insulating barrier between two metals which are in close contact. The quality of the tunnel barrier is expected to have a significant influence on the offset charge fluctuations~\cite{Zimmerman2008} and higher order tunneling processes~\cite{Pothier1994,Rajauria2008,Greibe2011,Maisi2011,Aref2011}, effects that ultimately limit the performance of single-electron devices.

In this letter, we present two complementary ways to study nominally identical tunnel barriers. Firstly, using transport methods, we characterize single-electron and two-electron tunneling through an aluminum oxide tunnel barrier between aluminum and copper. Measurements of the two tunneling processes allow us to estimate the homogeneity of the barrier. Secondly, using high resolution transmission electon microscopy (TEM) we image the cross-section of the tunnel junction barrier. The tunnel barriers in these two experiments were deposited simultaneously. Therefore they are expected to have similar characteristics. From the TEM images, we determine directly the distribution of the barrier thickness and demonstrate that these variations are a plausible explanation for the enhancement of higher order tunneling processes~\cite{Pothier1994,Greibe2011,Maisi2011,Aref2011}. The observed thickness variations are in line with independent studies of Ref.~\cite{Zeng2014}.

\begin{figure}[ht!]
	\centering
	\includegraphics[width=0.48\textwidth]{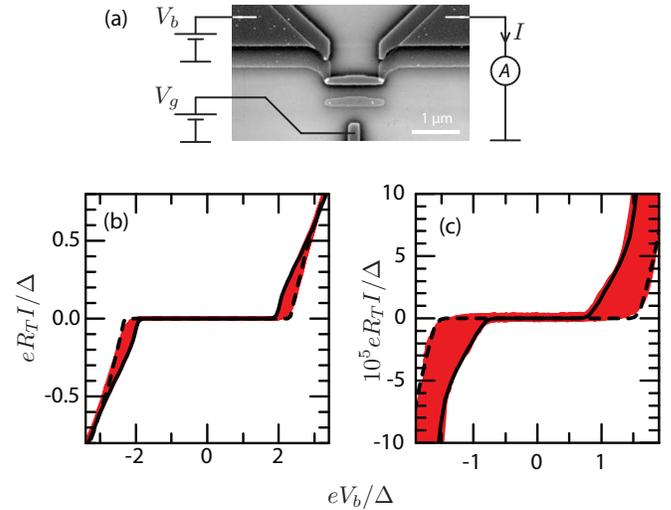}
	\caption{\label{fig:transport} (a) Scanning electron micrograph of the SET. Bias voltage $V_b$ and gate offset voltage $V_g = en_g/C_g$, where $C_g$ is the island-gate capacitance, are applied and current $I$ is measured. (b) Measured current $I$ as a function of $V_b$ shown as red colored region. The solid and dashed black lines are calculated at degeneracy ($n_g = 1/2$) and in Coulomb blockade ($n_g = 0$) respectively. On this scale the transport is determined by single-electron tunneling. (c) Subgap measurement similar to that of panel (b) but current scale zoomed in by a factor of about $10^4$. The current onset at $eV_b \approx 0.8 \Delta$ is due to Andreev tunneling.}
\end{figure}

The device in the transport measurements is shown in Fig.~\ref{fig:transport} (a). It is a single-electron transistor (SET) with superconducting aluminum leads and a normal metallic copper island. The SET is biased with voltage $V_b$ and current $I$ is measured. A gate voltage $V_g$ is applied to a gate electrode in order to vary the offset charge $n_g$ of the island. In Fig.~\ref{fig:transport} (b) we present measured current $I$ as a function of $V_b$ as a colored red region. The gate offset $n_g$ is swept over a full period of Coulomb oscillation. The measurement was performed at the $50\ \mathrm{mK}$ base temperature of a dilution refrigerator. Solid and dashed black lines represent numerical calculations for $n_g = 1/2$ and $n_g = 0$ respectively, based on single-electron tunneling. We determine the tunnel resistance $R_T = 75\ \mathrm{k\Omega}$ from the asymptotic slope, superconductor energy gap $\Delta = 210\ \mathrm{\mu eV}$ from the low bias regime where the current is suppressed, and the charging energy $E_c = 0.4\ \Delta$ from the Coulomb modulations. The area of the tunnel junction $A = 70\ \mathrm{nm} \times 120\ \mathrm{nm}$ is determined from the scanning electron micrograph of panel (a). The area $A$ and the tunnel resistance $R_T$ determine the transparency of the junction, $R_T A = 600\ \mathrm{\Omega \mu m^2}$, which is an essential parameter characterizing the tunnel barrier.

Figure~\ref{fig:transport} (c) presents a measurement similar to panel (b) but in the subgap regime $|eV_b|<2\Delta$ on a much smaller current scale. For a device with $E_c<\Delta$, the subgap current is dominated by two-electron Andreev tunneling~\cite{Averin2008,Maisi2011}, which has a threshold at $eV_b = \pm 2 E_c$. We determine an effective conduction channel size $A_\mathrm{ch} = 20\ \mathrm{nm^2}$ based on the slope of the I-V curve. The result is in agreement with previous findings and it is approximately an order of magnitude larger than the theoretically expected value $A_\mathrm{ch,e} \approx 2\ \mathrm{nm^2}$ giving rise to an order of magnitude enhanced Andreev tunneling~\cite{Pothier1994,Greibe2011,Maisi2011,Aref2011}. Next, we utilize TEM images of a barrier, deposited at the same process cycle as our SET, to demonstrate that the larger conduction channel area can be attributed to barrier thickness variations.

\begin{figure}[t]
	\centering
	\includegraphics[width=0.45\textwidth]{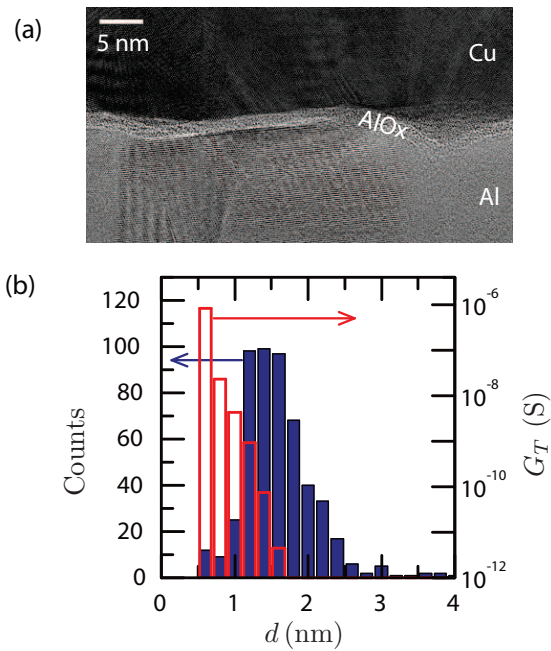}
	\caption{\label{fig:TEM} Transmission electron micrograph of an aluminum oxide tunnel barrier between aluminum and copper. (b), The distribution for the thickness $d$ of the aluminum oxide layer in solid blue bars and the resulting conductance $G_T$ for each bin as open red bars.}
\end{figure}

A typical cross-sectional TEM image of the barrier is shown in Fig.~\ref{fig:TEM} (a). The dark area on top is copper and the light area at the bottom is aluminum. The tunnel barrier in between consists of thermally grown aluminum oxide. The barrier thickness distribution is determined from many similar images covering a large sample area and the result is shown in Fig.~\ref{fig:TEM} (b). The thickness of the barrier varies from $0.5\ \mathrm{nm}$ to $4\ \mathrm{nm}$ with a mean value of $1.7\ \mathrm{nm}$.

We next calculate the conductance based on the thickness profile. We use a simple approach to divide the tunnel junction into several smaller areas and consider them separately as parallel tunnel junctions with uniform barrier thickness. The total conductance of the junction is obtained by summing the conductances of all areas. To model the conductance per area, we employ a model that describes tunneling through a trapezoidal potential barrier~\cite{Brinkman1970}. In our junctions, the asymmetric barrier profile is due to a difference in barrier height at the Al/AlOx and AlOx/Cu interfaces. Based on previous experiments~\cite{Shu1992}, we assume the barrier height difference to be $\delta \varphi = 0.25\ \mathrm{eV}$. The conductance per unit area of a tunnel junction at low bias voltage, $eV_b \ll \phi_\mathrm{max}$, and temperature, $k_BT \ll \phi_\mathrm{max}$, is
\begin{equation}
\label{eq:gTzerobias}
g_T = \left. \frac{dJ}{dV_b}\right|_{V_b=0} = \frac{me^2}{2\pi^2\hbar^3}\int_{-E_F}^0 d\epsilon_x P(\epsilon_x).
\end{equation}
Here $\phi_\mathrm{max}$ is the maximum height of the tunnel barrier, $J$ is the current density in the junction, $m$ the effective mass of the electron, and $E_F$ is the smaller Fermi energy of the two metals. $P(\epsilon_x)$ is the tunneling probability for transversal energy $\epsilon_x$, which according to the WKB approximation has the form
\begin{equation}
P(\epsilon_x) = \exp\left( - \frac{\sqrt{8m}}{\hbar} \int_{x_1}^{x_2} dx \sqrt{\phi(x)-\epsilon_x} \right),
\end{equation}
where $x_1$ and $x_2$ are the classical turning points, $\phi(x_i) -\epsilon_x=0$, and $\phi(x)$ the barrier height at position $x$. For an asymmetric barrier at low bias we have\cite{Brinkman1970}
\begin{equation}
\label{eq:barrierh}
\phi(x) = \bar{\varphi} + \left(\frac{x}{d}-\frac{1}{2}\right)\delta\varphi-\frac{1.15 e^2 \ln 2}{8 \pi \varepsilon \varepsilon_r d}\frac{1}{x(d-x)}.
\end{equation}
Here $\bar{\varphi}$ is the mean barrier height measured from the Fermi level and $d$ the thickness of the oxide. The second term of Eq.~(\ref{eq:barrierh}) corresponds to the slanting of the barrier. The third term describes the image forces which decrease the effective barrier thickness and lower the potential profile. In this term $\varepsilon_r$ is the dielectric constant of the oxide. We use a value $\varepsilon_r = 10$ for aluminum oxide. The results below remain the same, apart from a small adjustment of the fitted value of $\bar{\varphi}$, if different values of $\varepsilon_r$ and $\delta \varphi$ are used.

By using Eqs.~(\ref{eq:gTzerobias})-(\ref{eq:barrierh}) we determine the conductance $g_{T,i}$ for each observed thickness $d_i$ which we obtained from the TEM images. In Fig.~\ref{fig:TEM} (b) we plot the total conductance $\sum_i g_{T,i} A_i$ as open red bars, where $A_i = (1\ \mathrm{nm})^2$ is the area of each element. By summing over all elements and dividing by the total area, we obtain the tunnel conductance per unit area $G_T/A = (R_T A)^{-1}$. We fitted $\bar{\varphi} = 2.0\ \mathrm{eV}$ such that $R_T A$ matches the one obtained in the transport measurements. The free electron mass was used for $m$ in the calculations.

The conductance distribution of Fig.~\ref{fig:TEM} (b) demonstrates that the transport is strongly dominated by the thinnest parts of the barrier. The thicknesses with $d> 1\ \mathrm{nm}$ contribute less than $1\ \%$ to the total conductance despite that $95\ \%$ of the thickness values fall in this range. This finding already indicates that most of the tunnel barrier is inactive and suggests that thickness variations are a plausible explanation for the observations of large conduction channel area $A_\mathrm{ch}$. For a quantitative analysis, we write the ratio of the expected and observed conducting channel size as
\begin{equation}
\label{eq:ratio}
\frac{A_\mathrm{ch}}{A_\mathrm{ch,e}} = \frac{\Gamma_\mathrm{AR}}{\Gamma_\mathrm{AR,e}} = \frac{\sum_i G_{T,i}^2/A_i}{ G_T^2/A} = N \frac{\sum_i G_{T,i}^2}{(\sum_i G_{T,i})^2}.
\end{equation}
In the first equality, we used the fact that the Andreev tunneling rate scales as $\Gamma_\mathrm{AR} \propto A_\mathrm{ch}$. Subscript $e$ denotes the expected result by assuming a uniform tunnel barrier. In the second equality, we utilized the scaling $\Gamma_\mathrm{AR} \propto G_{T,i}^2/A_i$ for each element of the tunnel junction~\cite{Averin2008}. Here $G_{T,i} = g_{T,i} A_i$ is the conductance of the $i$th element. The total Andreev tunneling rate is obtained by summing over all sub-junctions. For the expected Andreev tunneling rate we have $\Gamma_\mathrm{AR} \propto G_T^2/A$, where $G_T = \sum_i G_{T,i}$ is the total conductance and $A = \sum_i A_i$ the total area of the junction. In the last equality, we assumed that the junction consists of $N$ pieces with equal area $A_i = A/N$. After plugging in the conductance distribution determined based on the TEM images, we obtain $A_\mathrm{ch}/A_\mathrm{ch,e} = 60$, using Eq.~(\ref{eq:ratio}), i.e., the Andreev tunneling is expected to be $60$ times higher in our junctions compared to a junction with a uniform tunnel barrier.

Our measurements indicate that barrier thickness variations give rise to large effective conduction channel area and hence to enhanced Andreev tunneling. However, the TEM analysis predicts by a factor of six larger enhancement than the transport measurements. There are a few possible explanations for this discrepancy. The TEM samples were prepared by annealing at $T = 80\ \mathrm{^\circ C}$. While this is not expected to change the characteristics, such as $R_T$ and the uniformity of the tunnel barrier, we cannot fully exclude this possibility. On the other hand, the TEM images were obtained from approximately $30\ \mathrm{nm}$ thick specimen. If the electron beam of the TEM is aligned with a straight barrier edge, we obtain a sharp image as in the center of Fig.~\ref{fig:TEM} (a). However, on the sides of the sharp section, the barrier is blurry which we attribute to the roughness of the barrier edge. From Fig.~\ref{fig:TEM} (a) we observe that the oxide layer is wriggling along the horizontal axis on this scale. We expect the barrier to wriggle along the transversal axis along the approximately $30\ \mathrm{nm}$ thick sample similarly. Such blurring contributes to an overestimation of the local barrier thickness, which leads to a larger value of $A_\mathrm{ch}$.

In conclusion, we have studied the uniformity of a tunnel barrier with two different approaches by combining transport measurements and TEM imaging. Both approaches indicate an enhancement of the effective conduction channel area $A_\mathrm{ch}$. The TEM analysis suggests larger increase of $A_\mathrm{ch}$ compared to the transport measurements. This discrepancy is attributed to the roughness of the tunnel barrier edges, causing an overestimation of the tunnel barrier thickness. Another possible, but less likely reason is the elevated temperature used during the preparation of the TEM specimen.

The work has been partially supported by Academy of Finland through the Centers of
Excellence Program (2012-2017) and the National Doctoral Programme in Nanoscience (NGS-NANO). We acknowledge the provision of facilities and technical support by Aalto University at Micronova Nanofabrication Centre and the Nanomicroscopy Center.


\bibliographystyle{aipnum4-1}
\bibliography{referencesurl}

\end{document}